\documentclass[sigconf,nonacm]{acmart}
\AtBeginDocument{%
  }

\newcommand{\secref}[1]{Section~\ref{#1}}
\begin{document}

\title{Echoes in the Digital Abyss: Examining the Bubble Surrounding
  Security and Privacy Discourse in Social Networks}

\author{Reagan Dennison}
\affiliation{%
  \institution{Northwestern University}
  \city{Evanston}
  \state{Illinois}
  \country{USA}
}

\author{Saanvi Sharma}
\affiliation{%
  \institution{Purdue University}
  \city{West Lafayette}
  \state{Indiana}
  \country{USA}
}

\author{Noshir Contractor}
\affiliation{%
  \institution{Northwestern University}
  \city{Evanston}
  \state{Illinois}
  \country{USA}
}

\author{Sruti Bhagavatula}
\affiliation{%
  \institution{Northwestern University}
  \city{Evanston}
  \state{Illinois}
  \country{USA}
}

\renewcommand{\shortauthors}{Dennison et al.}

\begin{abstract}
The dissemination of security and privacy education and guidance has been and still remains a challenge today.
Social networks represent a potential avenue for sharing best practices, and experimentally they have been found to be effective at this task.
While this appears promising, in the real world, security and privacy discussions would need to reach a wide range of people to be effective, avoiding the ``interest bubbles'' that commonly occur. We sought to understand how the communities surrounding security and privacy discourse operate, with a focus on what challenges need to be overcome to enable security and privacy discourse and advice to reach a wider audience.

Indeed, we found that in-the-wild security and privacy discussions in social
networks portray quite a different picture than in experimental settings. 
%
We built and analyzed the structure of a graph containing over 13 million users on the ``X'' platform (formerly
``Twitter''), 
including $10,159$ users who posted about security and privacy
and their followers.
Prior work has shown that users are more likely to consider information within social media if their like-minded social ties have visibly engaged with it. Our findings indicate that the users generating or participating in security discussions largely already belong to highly clustered technology and security- and privacy- related interest communities,
which suggests that the people who are not already in the ``inner circle'' of relevant interests
are likely not exposed effectively to these discussions.
We conclude with reflections and ideas on increasing the reach
of security and privacy guidance in social networks.

 
 

\end{abstract}




\maketitle

\section{Introduction}

With widespread technology use comes widespread risk of the technology being compromised, such that security and privacy concerns are ubiquitous.
This is evidenced by the rise
in the number of data breaches and security incidents
over the past few years~\cite{statistaNumberData}, with 2022 seeing around 1,800 data breaches~\cite{cnetDataBreaches}.
In an ideal world, the burden of security and privacy
is removed as much as possible from the user.
However, many systems do not prioritize security, treating it as an afterthought~\cite{gutmann2005security, steward2012software}. Additionally, the existence
of security threats that do not have clear technical solutions (e.g., social engineering attacks)
necessitate technology users to have awareness and exposure to security guidance and issues~\cite{awareness-behavior, awareness-behavior2}.

Prior work has highlighted the challenges with 
the dissemination of guidance related to security and privacy best 
practice and their acceptance~\cite{redmiles2016learned, redmiles2016think, rader2012stories}. 
Fortunately, social influence has been shown to have a significant impact
on users' security practices in experimental settings, including within social networks~\cite{das2014increasing, das2015role, emami2018influence, digioia2005social, rader2012stories, mendel2017susceptibility, avram2020engagement}. 
In the wild, however,
security and privacy discussions in social networks were found to
be too scarce to be likely to trigger social influence in the first place~\cite{bhagavatula2022adulthood}. Still, social media remains an effective channel
for information dissemination~\cite{bellini2021power, mueller2020saw, ye2020social} and so it is 
important to determine ways that security and privacy guidance or discussions can reach a wider audience
by exploiting the capabilities of social networks~\cite{weng2013virality}.

While the above is a longer-term goal, in this paper, we take an important step \textbf{towards} addressing this goal. To understand how to better propagate security
and privacy discourse on social media, we require an understanding of the network structure used to disseminate this information and what groups or communities are \emph{currently} generating and engaging with this discourse.
We hypothesized that
people more inclined towards technology and security and privacy already form a large contingent of the audience exposed to security
and privacy content, due to the nature of the platform (i.e., topic networks formed around common interests) and prior work~\cite{bhagavatula2021breach}. If these discussions are limited to ``interest bubbles'', or insular topic communities,
it would imply a barrier to entry for users who have a low awareness of cybersecurity issues or topics (and hence, are more likely to need the
exposure) to encounter guidance or information about security and privacy.

To that end, we are interested in answering the question: \emph{Are security and privacy discussions in social networks limited to interest bubbles?}  

In order to determine whether these communities are both insular and limited to users with pre-existing related interests, we studied two research questions that tackle the above broad question
in different ways: 
\begin{enumerate}
    \item Are security and privacy discussions primarily occurring among users in close-knit network communities?
    \item Do the users who engage with (i.e., create, retweet, quote, or reply to)
    posts about security and privacy display common interests and pre-existing inclinations towards security and privacy?
\end{enumerate}


To study both these questions,
we analyzed a set of 10,159 users who interacted with security and privacy content (we will call these users ``interactors''), 
on the ``Twitter'' social network
platform (currently known as ``X''; we discuss the relevance of using this platform in Section~\ref{sec:discussion})\footnote{We will refer to the ``X'' platform as ``Twitter'' throughout this paper as it was still ``Twitter'' at the time of data collection.}. For the first question, we built a graph consisting of the 
6,960 interactors and their followers, resulting in a graph
with 13.74 million user nodes; we analyzed its communities. For the second question, we separately analyzed posts that each interactor interacted with as a proxy for their ``interests''. We built topic models over each users' interests to assess the degree of shared interests.

Our results from both analyses indicated that security and privacy discussions in our dataset were occurring in close-knit communities of like-minded people with security-related interests. 
This suggests that the users that would benefit from the exposure to security and privacy discussions may not actually be the ones to have much meaningful exposure to this content through their social networks. 
This implication is in line with what we know to be true
of social networks in that users are recommended what they're interested in.
However, the uniqueness of
security and privacy as a topic lies in its universal relevance to
every individual utilizing technology. Our findings highlight
the specific problem that needs to be overcome for security and privacy discussions guidance to reach beyond communities with homogeneous interests.


\section{Related work}
\label{sec:rel_work}

We survey three categories of related work: (1)~exposure to security and privacy information; (2)~security and discussions in social networks;
and (3)~domain-specific social network analysis.

\subsection{Security and privacy exposure}
\label{sec:rel_awareness}

Research has shown that exposure to security and privacy news has a beneficial impact on individuals' security and privacy literacy ~\cite{morrow2022newsconsumption}.
Much of previous work related to security and privacy exposure has been studied in the context
of data breaches and security incidents. One such study found that only a small portion (16\%)
of users were aware of widespread breaches for example, the Equifax or Yahoo! breaches~\cite{bhagavatula2021breach}.
They found that this percentage was low even when users were affected by the breaches. This finding is also supported by another recent study~\cite{mayer2021now} that found that participants were unaware of 74\% breaches that they were presented with. Another study conducted by the Rand corporation found that 
almost half of the users they surveyed learned about a data breach from a source other than the company that was affected, indicating that this notification was likely not easily available~\cite{ablon2016consumer}.
Other existing work studied generally, the various ways in which people learned about breaches and found that social media played a major role in informing users, accounting for almost a third of the sources reported~\cite{das2018typology}.

Prior studies have also examined who is receiving security and privacy guidance or information on best practices.
For example, it was found that users in lower socio-economic classes were less likely to receive security and privacy advice,
indicating a digital divide in the dissemination of this information~\cite{redmiles2016learned}. One of the aforementioned studies
also found that the people who were likely to be exposed to news about a breach were ones that browsed technology-related articles,
displaying an inclination that may have influenced the information they came across~\cite{bhagavatula2021breach}.

Our work is motivated by related works'
findings that 
cybersecurity and privacy exposure is limited among computer users
and is skewed to users in certain groups.
It is additionally motivated by the prevalence of social media as an information source.



\subsection{Security and privacy discussions in social networks}
\label{sec:rel_discussions}

Social influence has been found to be effective in promoting security
and there have been a number of studies that have espoused the usefulness of
social networks in implementing this.

Prior work has shown that social media is an effective channel for increasing awareness of breaches~\cite{das2018typology}. Moreover, research reveals that Twitter serves as a forum for individuals to express their viewpoints on security incidents and utilize the platform as a means to disseminate guidance to others~\cite{dunphy2015socialmedia} 

Experimental studies have found that when users in a social network shared security practices they implement,
their friends were also likely to adopt those features~\cite{das2014increasing, das2015role, digioia2005social}. However, work studying security and privacy discussions
in the wild found that there is a low volume of security and privacy discussions on social networks, likely insufficient to trigger the same type of influence~\cite{bhagavatula2022adulthood}.

While most of these studies suggest social media as an effective tool
for encouraging constructive security practices, one of them suggests that there is work to be
done to increase the engagement with security- and privacy-related content
for social networks to be an effective medium. Our work attempts to identify gaps in engagement for which future work can aim to fill.


\subsection{Domain-specific social network analysis}

Given the usefulness of social networks in information dissemination and diffusion, they have been
prominently studied in the context of a variety of domains. For example, studies on social networks have observed patterns of happiness, monitored university students "fear of missing out", and disaster response~\cite{dodds2011happiness, li2022fomo}.
Prior work demonstrates how applying these tools provides insight into the way individuals maintain topic networks ~\cite{wasim2020covid5g}, and recent work utilizes social media data to map out user ties within these topic networks~\cite{benigni2017isis, yao2021construction}. 

In recent years, Twitter has emerged as a prominent resource for studying social network structures, given its textual and interaction modality information ~\cite{miller2011twitter}. Prior research indicates that non-viral information within Twitter social network structures spreads as complex contagion, such that the most effective exposure to information comes from social reinforcement within communities with high homophily. ~\cite{weng2013virality}
In general, in social networks it is not uncommon for users to be exposed to content they're already interested in, like sports or political networks \cite{pariser2011filter}. However, given that security and privacy concerns are ubiquitous to every computer user, the security and privacy social network space is more similar to that of other topics of public interest, like weather events or public health messaging.
For example, a study from 2016 investigated how opinions about the HPV vaccine clustered on Twitter, by conducting community detection on a follower network and a textual analysis of tweets, providing a way to characterize these communities so that further methods can address groups with less exposure to accurate information ~\cite{surian2016hpvvaccines}. Similarly a study from 2013 explored how disaster response misinformation propagated in the wake of Hurricane Sandy in order to gain a better understanding of the network topology that allows this information to spread~\cite{gupta2013hurricanesandy}.
While these works are similar to our work in respect to universally-applicable topics,
additional recent works have used following networks within Twitter to understand information propagation within networks~\cite{beguerisse2014ukriots, benigni2017isis}.

While previous work in the field of security and privacy uses social network analysis to measure software vulnerability awareness \cite{shrestha2020softwarevulnerability}, there is little work in studying the network structures used to propagate security
and privacy information. Previous work on social network structures indicate that networks with greater number of edges and lower numbers of singletons disseminate information rapidly ~\cite{romero2013topicalstructure}. In this paper, we
bring the problem of security and privacy information dissemination (studied experimentally
as described in \secref{sec:rel_discussions} to real
social network structures, drawing inspiration from the aforementioned studies.
Specifically, using social networks, we measure the state of communities that are exposed to security
and privacy engagement.

\section{Data collection}
\label{sec:data}

We collected and built our data set in steps between 
July 2022 and March 2023. The first step was compiling the interactors
(\secref{sec:interactors}); this set of users served as the basis for addressing our two research questions. To answer our first research question, we collected data about the interactors to build a follower network
modeling the connections between
these interactors and their followers (see \secref{sec:graph}). To study
our second research question, we collected data
representing the types of content each interactor interacted with
(\secref{sec:interests}). 


\subsection{Collecting security- and privacy-related posts}
\label{sec:interactors}

To curate a set of interactors of security and privacy content,
we needed to start by identifying a set of security and privacy
posts and then extract the users that created those posts, or retweeted, replied to, or quoted them.

Ultimately, we collected this set of posts related to security and privacy
by searching for Twitter posts against a set of terms related to cybersecurity and privacy (``key terms'', henceforth). To build this set of key terms, we first created an initial set of key terms by searching for ``cyber security'' and ``digital privacy'' on Twitter (specifically, posts in English) using the API~\cite{twitter_api} and manually examining the ``top''\footnote{The ``top'' posts are the default ones returned by searching.} 100 posts
returned for each of these two search queries.
We examined each post and extracted keywords or terms within these posts,
which we believed would produce content relevant to cybersecurity and digital privacy,
when searched for.
This process produced 168 key terms.

Not all key terms reliably produced Twitter posts relevant for security and privacy (e.g., ``vulnerabilities'' can return posts about personal vulnerabilities, ``insider threat'' can produce posts that contain those two words but do not refer to cybersecurity threats).
Therefore, using an
approach outlined in prior work~\cite{bhagavatula2021measuring},
we reduced this set of key terms to ones we could be confident would return relevant posts, when searched for.
Specifically, we evaluated which key terms to filter out in the following way:
we searched for each of the key terms on Twitter and collected the top 20 returned posts.
On manual observation of each of these 20 posts,
if we observed all of these posts were related to cybersecurity or digital privacy, we included the key term in our final set of key terms.
If the the number of relevant posts was between 10 and 19,
we included the key term in our final set but when the posts were later collected
against these terms (described below), we would filter them for true positives.
If the number of relevant posts was less than 10, we did not include the key term
in our final set.
After
this filtering process, we were left with 81 final key terms. Table~\ref{tab:keyterms} in Appendix~\ref{app:key_terms} contains the key terms we considered, indicating
which we kept, which we filtered, and which we discarded. 
Through this process, a few important terms such as ``passwords'' were not included as key terms given that they appeared in non-technical or irrelevant contexts too often for us to include it or include but filter it (i.e., it met the third criteria above). However such terms appeared often in posts fetched against the other included key terms. Therefore, we believe this final list of terms to be sufficiently relevant and inclusive.

We then collected up to 200 posts for each of the final key terms 
and removed any duplicate posts (we kept retweets if the retweet of an original post was retweeted by a separate user). We manually filtered posts that were collected against key terms which met the second criteria above. This resulted in a set of $8,268$ posts. Table~\ref{tab:example_posts} in Appendix~\ref{sec:example-posts}
has a few examples of these posts. We evaluated its quality by manually observing a random sample of 50 posts in the same way as before; we observed a $90$\% true positive rate.

\subsection{Extracting interactors}
\label{sec:graph}
From the 8,268 posts collected in \secref{sec:interactors}, we extracted all the Twitter
users who were responsible
for any of the tweets in the set, whether the tweet was an original, a reply, a retweet, or a quoted retweet. 
If the tweet was a reply, quote, or retweet, we also collected the user that authored the original post. This resulted in a list of 6,960 interactors, to start with.

For each such user, we then collected their follower count using the Twitter API~\cite{twitter_api}.
Given that the Twitter API rate-limited requests for follower lists to 75,000 follower IDs per request, we first collected all the follower lists for users with under 75,000 followers. This resulted in 13.27 million additional users collected through these follower lists. We reduced this set by ensuring that followers that were shared among more than one interactor or followers that were already interactors were only counted once. 
For interactors with over 75,000 followers, while data collection was difficult, we
were also concerned about including their followers if their content
was diluted with a high volume of content unrelated to security and privacy (e.g., accounts of popular news sites). This is because
we believed that followers of such accounts had a low chance of encountering the very little security-related content the large account posted.
Instead, we included a large account
if they posted about security and privacy often.
We decided this by collecting 20 of each of these interactors' most recent tweets
and manually examined them to see if the posts were related to cybersecurity or digital privacy.
If we found at least 5 relevant tweets, we collected their follower lists; this applied to 36 out of 311 users with large follower lists.
We still kept the remaining 275 users with at least 75,000 followers in our dataset, without collecting their followers.



We built a graph containing all of the above collected users as nodes. 
We were unable to collect the followers for 151 interactors who were suspended or whose accounts were private at the time of follower collection. These users were still added to the graph as nodes, but their followers could not be collected.
We then collected additional data via the API to identify existing users in the graph who retweeted any of the original $8,268$ posts. Adding this additional data, $3,199$ users were identified as interactors and added
to our original total of $6,960$ for a \textbf{new total of $10,159$ (no new nodes were added as these new interactors were already followers initially present in our network)}. However, we did not collect the followers of these additional retweeters
since they were not returned as interactors in our original data collection. \emph{We henceforth refer to this augmented set as the ``interactors'' when referring to the graph analyses.}



\subsection{Collecting data for topic models}
\label{sec:interests}
To perform topic modeling analysis on the interactors' ``interests'', we needed a way to represent these. 
We used the kinds of posts that the interactors interacted with in general
as a proxy for their interests. For each interactor, 
we collected up to $20$ of the latest posts that each of them
interacted with (these posts were collected
from their profile,
separately from their security-related posts). 
This resulted in $13,393$ tweets across $6,960$ users. After removing duplicate tweets (i.e., when we collected an original post and retweet),
$9,766$ tweets remained to analyze.



\section{Methodology}
\label{sec:methodology}

We describe our approaches to both research questions in this section.
We addressed the first research question
by modeling a subset of
the Twitter social network graph seeded from posts relevant to cybersecurity and digital privacy (\secref{sec:meth_graph}). 
We addressed our second research question via a
textual topic model analysis on the ``interests'' of interactors (\secref{sec:meth_text}).


\subsection{Social network graph analysis}
\label{sec:meth_graph}


We built a follower network as a weighted directed graph for the first research question. The purpose of this graph
was to model the relationships between users who actively propagated security and privacy information (i.e., the interactors) and their followers. We then used community detection algorithms to partition the graph into communities, and measured both global properties of the graph as well as local properties of each community to determine how far the reach of security and privacy information extends.


\subsubsection{Building the graph}
\label{sec:meth_building}

Each node was a user we collected
in \secref{sec:graph}. A directed edge existed between two nodes if one user 
followed another on Twitter. Table~\ref{tab:graph_components} describes the different graph components (some described below) and how frequently they occurred in the graph.


We hypothesized that the specific ways users interacted with the original security and privacy posts (if applicable) represented
a different contribution and influence to the propagation of information. 
In particular, we differentiated between users that produced some content by posting, quoting, or replying (we will call them ``authors'') and those who just passed along, boosted, or endorsed information by retweeting; this is supported by prior work~\cite{tornberg2018echochamber, surian2016hpvvaccines}. 
Therefore, we assigned additional attributes to interactor nodes to indicate whether they were an
author and/or a retweeter.
We also highlighted connections in the graph that could have led to a retweeter
retweeting an author's post (whether or not the retweeter was a follower of the author). Hence,
we weighted each edge between two nodes based on the role the edge might have played in
connecting retweeters to the authors of security and privacy posts they had retweeted.

\begin{table}[h]
\begin{tabular}{|p{5cm}|p{3cm}|}
\hline
\textbf{Graph component} & \textbf{Frequency} \\ \hline
\hline
\#interactors                            & 10,159 \\ \hline
\#user nodes in the graph                 & 13,740,737 \\
\hline
\#edges from a follower to an interactor & 23,613,084 \\
\hline
\#retweeters with paths to the author they retweeted    & 2,838 \\ \hline 
\end{tabular}
\caption{The various components included in the graph
and how often they occurred. Recall from Section~\ref{sec:graph} that the number of interactors increased from $6,960$ to $10,159$.} 
\label{tab:graph_components}
\end{table}



We computed weights for edges to assign more importance to edges that could
have been responsible for influencing a user to passively interact with content
(i.e., by retweeting).
A retweeter and an author could have been connected by one edge or a path consisting of multiple edges. If there were multiple paths from a retweeter to the author they retweeted, we only considered the 
shortest path.
For an edge that belonged to a path between a retweeter and the author they retweeted, we increased that edge weight as follows.
We calculated the shortest path from each retweeter to the author they retweeted, if such a path existed. 
We associated each such path with a value, which was meant
to represent the strength of the author's potential influence
in leading to the retweet.
We calculated this value for a path by dividing the number of the author's tweets which the user retweeted, by the number of edges in the path. This definition assigns a higher value to paths that could have been responsible for more retweets, and a lower value for longer paths, indicating that the author was less likely to be responsible for the retweets. 
All edges started with a weight of zero. For every edge that appeared in one path, the path value was added to the edge's existing weight. For an edge that appeared in more than one path, the path values of all the paths it belonged to were added up to produce the edge's weight. For an edge that appeared on no paths, its weight remained as zero.


Formally, we define the weight of one edge as follows: 
\[
W_e = \sum_{i=1}^{|P|} \frac{r_i}{l_i}
\]

\begin{figure}[h]  
  \begin{minipage}[b]{0.48\linewidth}
    \centering
    \label{fig:all}
    \includegraphics[width=.5\linewidth]{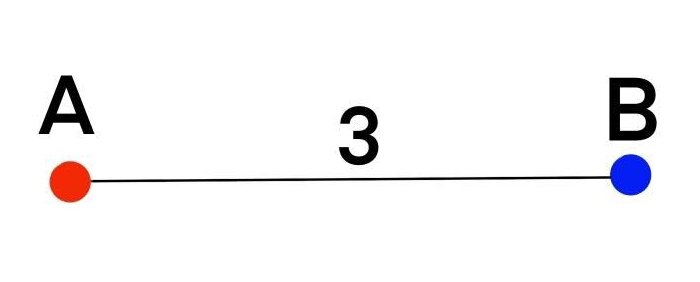}
    \caption{User A retweeted three of author User B's posts. There is one direct edge between them, and so $\frac{3}{1}$ is added to the edge's weight (which was initially 0).}
    \vspace{4ex}
    \label{fig:direct-edge}
  \end{minipage}
  \hfill{}
    \begin{minipage}[b]{0.5\linewidth}
    \centering
    \includegraphics[width=.5\linewidth]{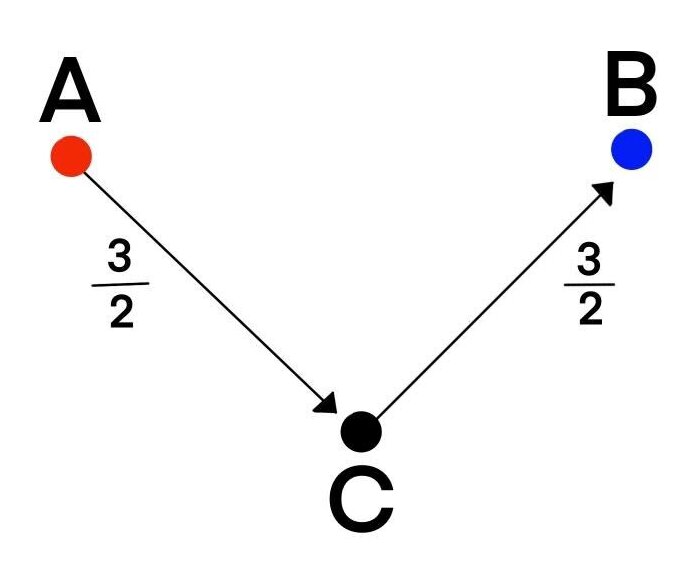}
    \caption{User A retweeted three of author User B's posts. The path from User A to User B, goes through User C who is not an author.
    There are two edges connecting User A to User B and each edge only occurs on this one path.
    Therefore, $\frac{3}{2}$ is added to each edge weight (which was initially 0).}
    \vspace{4ex}
     \label{fig:indirect-edge}
  \end{minipage}
  \begin{minipage}[b]{\linewidth}
    \centering
    \includegraphics[width=.25\linewidth]{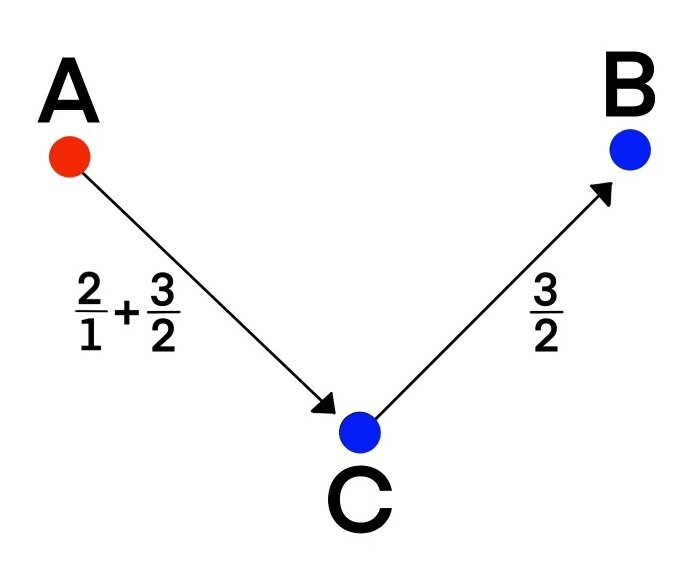}
    \caption{User A retweeted three of author User B's posts and two of author User C's posts. The edge from A to C appears in two different paths that connect a retweeter (A) to an author (C or B).
    The edge C->B only appears in one path (A->C->B), therefore, $\frac{3}{2}$ was added to its weight (initially 0). 
    The edge A->C appears in both the A->C path and the A->C->B path, therefore, $\frac{2}{1}$ and $\frac{3}{2}$ are both added to its weight (initially 0).}
    \vspace{4ex}
    \label{fig:indirect-edge2}
  \end{minipage} 
\end{figure}

Here, $W_e$ represents the weight of a single edge. $P$ represents the set of all paths to which the edge belongs and $i$ represents
an index into this set.
For one path $P_i$, we define $r_i$ as the number of times the user at the start of the path retweeted the author at its end. We define $l_i$
as the number of edges on the path.
Figures~\ref{fig:direct-edge},~\ref{fig:indirect-edge}, and ~\ref{fig:indirect-edge2} 
provide examples of how the weight for an edge was computed in a few different cases.

\subsubsection{Measuring characteristics of the graph}
After we constructed the full graph, the next step
was to analyze how the interactors were distributed across
the network or if there were notable ways in which
users in the graph were grouped.

We used Louvain’s community detection algorithm~\cite{blondel2008fast} implemented as part of the Python NetworkX library~\cite{networkx_api} to partition the graph into communities. This required treating the graph as an undirected graph. We also measured other global properties of the graph such as modularity~\cite{newman2006modularity} and density~\cite{anderson1999density}. Modularity is widely used to determine the optimal division of a network into communities, indicating if groups of nodes tend to cluster together with few external connections. The graph modularity can shed light on whether communities are highly isolated, indicating if ties between communities are weak and/or nonexistent, or strong. Density measures the level of connectivity in a graph by assessing its closeness to its maximum potential connectivity.

After identifying communities,
we focused on communities with more than a single user.
Within these communities, we observed the percentage of all interactors (identified in \secref{sec:graph}) represented in these communities, as well as what percentage of a community itself was composed of interactors. These observations enable us to determine whether or not security and privacy discussions
are siloed within certain communities.

We discuss these findings in \secref{sec:res_graph}.


\subsection{Textual analysis of interactors' interests}
\label{sec:meth_text}

\subsubsection{Topic Modeling}

Analyzing the graph as we did in \secref{sec:meth_graph}
produces quantitative results, highlighting how isolated security and privacy discussions can be within
a social network. From these findings, we can speculate
whether or not the communities formed
consist of like-minded people who are already inclined to security and privacy, explaining their connections.
However, we can further discern that via a more qualitative analysis.

We supported the graph analysis with a qualitative analysis of the ``interests'' of
the original $6,960$ interactors. We built topic models over these
``interests'' to identify any underlying common topics that these users had an inclination for.

We considered the types of content an interactor interacted with on Twitter as a proxy for ``interests''
as people tend to interact with content aligned with their interests.
To that end, we determined the ``interests'' of an author
from the 20 posts they interacted with at the time
of data collection. We tokenized all the posts
and represented one interactor's interests as a bag-of-words collection of
the tokens from these 20 posts.

In the context of topic modeling, we treated each user's bag-of-words collection as one document,
after retaining only nouns~\cite{martin2015more} and removing stopwords.
We computed term frequency scores and TF-IDF scores for each token in the document~\cite{DBLP:books/daglib/0021593}, which would serve as
the inputs to the topic modeling algorithm.

We experimented with topic models using both the Non-negative Matrix Factorization (NMF) or Latent
Dirichlet Allocation (LDA) algorithms~\cite{lee1999learning, blei2003latent}
and determined which model produced the most coherent output.
For each algorithm, we identified the optimal number of topics by varying 
the number of topics between two and 10, and for each, computing the intra-topic similarity as seen in prior
work~\cite{bhagavatula2021breach}. We constructed our topic models
using each algorithm with that optimal number of topics.

For the model produced by each algorithm, three members of the research team manually observed the 30
most frequently occurring tokens in each of the resulting topics
so that we could label them based on overarching themes. After manual observation, they labeled the topics by each of them proposing possible labels and finally converging in agreement on a label.

We hypothesized that the interactors of security- and privacy-related content
displayed a common interests or a shared inclination to security and privacy.
If we identified only a few coherent topics displaying these interests, that would support our hypothesis.
On the other hand, it would not be supported if we were not able to discern any coherent topics or
a high number of topics.

\section{Results}
\label{sec:results}

In this section, we describe the results of the graph analysis from \secref{sec:meth_graph} and the textual analysis
of the interactors' interests in \secref{sec:meth_text}
to address our two research questions.


\subsection{Social network graph analysis}
\label{sec:res_graph}

\subsubsection{Global graph statistics}
The graph we built as described in \secref{sec:meth_building}
contained 13.74 million nodes, with 10,159 of those nodes representing interactors. 
In total, interactors made up 0.07\% of the graphs’ nodes, illustrating that the vast majority of their followers had no measured engagement with the original set of posts. The graph showed a density of $1.25x10^{-7}$, indicating that the 
graph was sparse, which was expected given the large number of non-interacting nodes in the graph as a result of collecting 
all the followers of each interactor. 

We computed a modularity of $0.801$ on our graph, which implies that the graph
was strongly divided into communities with dense inner connections and few external connections to other communities. 
Given that social network communities can "trap" information and prevent it from spreading further due to their overarching network topology, as well as due to limited user attention~\cite{weng2012attention}, we sought to understand these communities through community detection,
whose results we describe in \secref{sec:res_community}.

\subsubsection{Community detection results}
\label{sec:res_community}
The high modularity motivated us to identify and analyze the communities to glean
whether interactors tended to group together. 
Louvain's algorithm identified 13.7 million total communities; this includes communities that just contained a single user node.
The high single community count was not surprising given that the majority of the graph nodes belonged to the followers of the original interactors
who were not interactors themselves.

For the purpose of studying the community structure in which security and privacy discourse occur, we did not consider such single-user communities as significant in terms of network topology and removed them from our analysis,
leaving us with 176 communities. 
Despite making up 0.07\% of the total nodes in the graph, the interactors in the graph
accounted for 99.89\% of the users in the non-single-user communities. 
22.27\% of all the $10,159$ interactors ($2,262$ interactors) were part of the non-single-user communities.

While we found several small communities, we were interested in the larger communities
as we believed they would give us better insight into how interactors fit into
communities. The eight largest communities each contained over 50 users. There were no non-interacting users in the these largest communities. 
Table~\ref{tab:communities} summarizes the properties of 
these eight largest communities.


\begin{table}[]
\begin{tabular}{|p{0.3\linewidth}|p{0.3\linewidth}|p{0.3\linewidth}|}
\hline
Community size & \% of community nodes that are interactors & \% of interactors in this community. \\ \hline
\hline
501            & 100\%                                      & 22.15\%                              \\
343            & 100\%                                      & 15.16\%                              \\
282            & 100\%                                      & 12.47\%                              \\
160            & 100\%                                      & 7.07\%                               \\
153            & 100\%                                      & 6.76\%                               \\
84             & 100\%                                      & 3.71\%                               \\
75             & 100\%                                      & 3.31\%                               \\
58             & 100\%                                      & 2.56\%     \\ \hline                          
\end{tabular}
\caption{Information about each of the eight largest communities, including the number of nodes in the community, the percentage of that number that are interactors, and the percentage of total interactors that are present in the community. For the third column, the denominator for the percentage was the $2,262$ interactors in non-single-user communities.}
\label{tab:communities}
\end{table}


The high modularity value within the non-single communities suggests that these smaller groups of users engaged in security discussions tend to remain somewhat separate from other communities and users. Combined with the disproportionately high representation of interactors in these non-single communities, our results indicate that those who engage with and boost security and privacy discourse primarily do so to other users who already actively contribute to these conversations, meaning that these discussions are unlikely to reach outside of these close-knit communities.

\subsubsection{Example Community}
\label{sec:res-comm-5}
\begin{figure}[ht]
    \centering
    \includegraphics[width=0.4\linewidth]{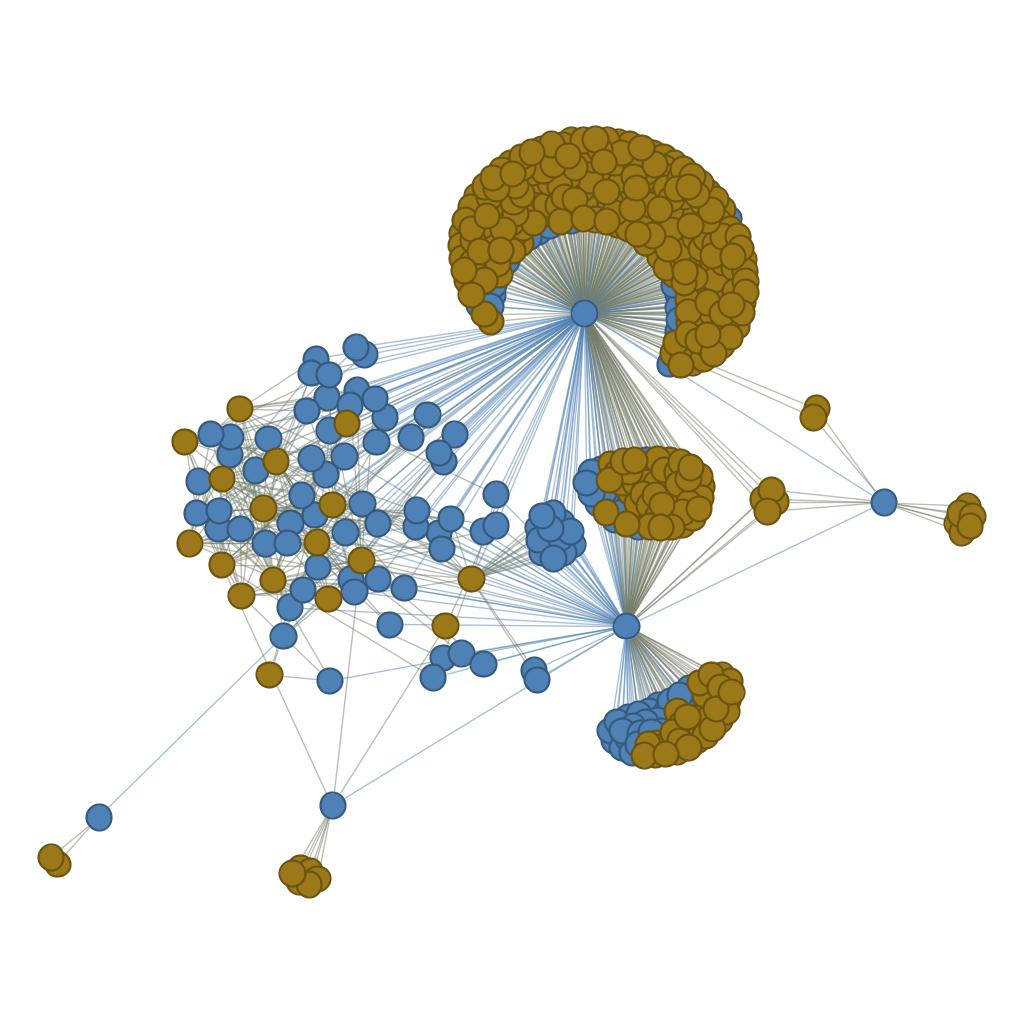}
    \caption{Visualization of one of the largest communities, with a size of 153 users. Interactors part of the community are colored in blue, while interactors outside of the community are colored in yellow. In this particular community, every user is an interactor, and follows other interactors. In fact, 88.5\% of community members are authors of security and privacy related content.}
    \label{fig:comm-visualization}
\end{figure}

To gather more insight into the community structures, we visualized one of the largest communities with 153 users
in Figure \ref{fig:comm-visualization}.
Similar to most of the communities, this community was entirely comprised of interactors.
Furthermore, 88.5\% of users in this community were authors, with only 11.5\% as non-authors (only interactors). This high concentration of authors indicates this community especially contributes to security and privacy discussions, and their social network serves as the first avenues these discussions must travel through to reach a larger audience. As illustrated in the figure, there are 'hub' users that connect to other interactors outside of the community whom rely on this following relationship in receiving discussions from inside the community. The high prevalence of interactors within this community points to a tendency for interactors to form into tight-knit, insular communities.

\subsection{Textual analysis of interactors' interests}
\label{sec:res_text}

We computed topic models using the NMF algorithm~\cite{lee1999learning} and the LDA
algorithm~\cite{blei2003latent}
and found that NMF produced more coherent topics (also supported by prior work~\cite{egger2022topic}).
As such, the following results are based on the results of NMF topic models.

We found that the within-topic similarity was highest when the number of topics was three.
We observed the 30 most frequently occurring tokens in each of these topics
and labeled each topic if one was evident. Table~\ref{tab:topics} shows
the top 30 keywords for each of the resulting topics.
Through this process, we were able to label two of the three coherent topics. Table~\ref{tab:topics} also shows the label that we gave to those two topics,
which generally indicated an interest in general technology and security and privacy. The one incoherent topic suggested the presence of other interests, which was to be expected, but the two coherent topics still display a pattern. 
We sanity-checked these results to ensure that the topics we found reasonably encompassed most of the content we collected about each interactor; we describe how we verified this in \secref{res:comprehensive}.

These findings answer our second research question in the affirmative.
They suggest that interactors may have shared interests
and do demonstrate a pre-existing inclination towards technology or security and privacy already. This serves as additional reinforcement that many of the users who engage in security and privacy discussions do so within shared interest bubbles.


\begin{table}[t]
\begin{tabular}{|p{0.80\linewidth}|p{0.15\linewidth}|}
\hline
\textbf{Top 30 tokens} & \textbf{Label}\\ \hline
\hline
cybersecurity, security, infosec, ransomware, threat, attack, blog, attacks, risk, cloud, tools, data, team, information, report, systems, services, business, details, post, today, management, system, industry, service, technology, solutions, top, software, week & Security and privacy
 \\ \hline
data, crypto, ai, privacy, business, tech, project, technology, market, cloud, platform, future, services, software, solutions, management, code, industry, community, companies, team, information, july, article, experience, research, company, key, top, service & General technology
 \\ \hline
like, people, one, time, day, years, right, today, first, year, life, way, world, two, state, twitter, money, things, support, media, health, days, something, government, thing, week, home, change, end, power & Incoherent
 \\ \hline
\end{tabular}
\caption{Each row in this table represents the top 30 tokens for one topic as well as the label
we assigned to the topic. A label of ``Incoherent'' means we were not able to identify a clear topic.}
\label{tab:topics}
\end{table}

\subsubsection{Validating topic comprehensiveness}
\label{res:comprehensive}

We needed to ensure that we did not just miss seeing other interests 
because we only included tokens that appeared in at least 10\% of the documents (also known as the document frequency cut-off).
To verify this, we manually examined all the discarded tokens that we did not use in the model.
The discarded tokens mostly consisted of usernames or hashtags such as ``\#dataliteracy'', 
``\#efficacytesting'', ``passwordcracking'', ``rulesofdigitalmarketing'', and ``secretsofcrypto'' among them. The discarded tokens did not
indicate any added discernible interest and they did not occur frequently enough to make a difference 
in the model results if we included them.
\section{Limitations}
\label{sec:limitations}

Our dataset is subject to a few limitations. Even though we analyzed over $8,000$ tweets, this can be considered small when compared to the size of Twitter. However, the limitations of the Twitter API at the time made it difficult to collect more than $200$ posts per key term. Furthermore, we believe that the dataset is sufficiently large (over $8,000$ posts) to represent a meaningful cross-section of the users of interest.

Our dataset is also likely biased if Twitter returned posts specific to our geographic location. However, we believe it is unlikely for results from other parts of the country where we reside (USA) to be significantly different from each other based on regional location.

Additionally, due to the large number of nodes we ended up collecting, we were not able to add in ties that might have existed between collected followers beyond the ties to the original interactors. However, prior work supports our approach of seeding a network from posts about a known topic and building a follower network to study the network structures surrounding those topic discussions~\cite{benigni2017isis, yao2021construction}.

We were not able to collect follower data for all users with over 75,000 followers due to rate limits imposed by the Twitter API; we were only able to collect the followers of 36 out of 311 users. We don't believe that including the remaining users' followers would
have impacted the big picture of our findings, given that the users we did not include likely did not have a large following interested in their security and privacy content. By not including their followers, we ensured that we were not considering large accounts such as news organizations as security and privacy interactors simply because of a single post they made. 

We also could not collect the followers of 151 original interactors because their user accounts switched to private or were suspended by the time we started collecting followers. However, these interactors for whom we could not collect followers only accounted for
2.17\% of the original 6,960 interactors, and therefore, including their followers would have been unlikely to alter
our findings and implications significantly.

We discuss ``exposure'' to content about security and privacy but only measured exposure by interactions with the original
set of posts about security and privacy. Realistically, a user can come across content while scrolling, read it and then moving on without interacting with it.
There is no guarantee, however, that in this case the user has understood or processed the post
especially if they encountered it during idle scrolling. We were specifically interested in users
who actively engaged with content which was evidenced by their interactions. Previous research indicates that visible engagement from those in a user's network means that users are more likely to consider said information as valuable, and worth further sharing ~\cite{avram2020engagement}. Moreover,
prior work has established the usefulness of recording engagement through
explicit retweets~\cite{tornberg2018echochamber}, which we considered as well.

As part of building out paths from
an author to a retweeter, we could only determine a path from 36\% of the retweeters to all authors for whom they retweeted. Given that our graph is based off of the original interactor's followers, we may have missed out on following relationships to connect retweeters to authors.
The absence of a straightforward path
from a retweeter to the author they retweeted may also have been due to algorithmic recommendations that were difficult to trace.

Finally, a large percentage of the interactors were part of single-user communities, which we did not consider as communities. This was likely due to those users not having a large following
who engaged with their security content. 
\section{Discussion}
\label{sec:discussion}

Analyzing the data of $10,159$ Twitter users, we attempted to answer our original research questions
of whether security and privacy discussions occur in social media bubbles
and whether users who interact in these discussions share characteristics.
Regarding our first research question, building a follower network graph stemming from these $10,159$ users, we found eight distinct large communities, all consisting of users who actively engaged with security- and privacy-related content. This finding was surprising given that we collected the original $6,960$ interactors independently of each other and we did not assume there to be any connections between them. Regarding our second research question, we analyzed the interests of interacting users
by looking at the types of posts they generally engaged with on Twitter and found that the general interests of these users already largely aligned with either security and privacy or general technology. 

We determined one main takeaway from our findings: homogeneity in interests combined with the insular nature of the communities indicates the information may not be sufficiently reaching audiences outside of a technical and security and privacy oriented sphere, i.e., the people that do not typically look at this information.

This takeaway gives rise to two challenges to overcome: (1)~increasing the reach of security and privacy guidance
or increasing the participation in security and privacy discussions; and
(2)~increasing the acceptance of security and privacy guidance. We next discuss the implications of our findings and recommendations addressing these two challenges. We discuss these below along with a discussion about studying ``Twitter''.

\subsection{Increasing the reach of security information}

Our findings suggest
that for security and privacy discussions to reach wider audiences, the information needs to have an influence outside of close-knit security-inclined
communities, i.e., it needs to jump across social network communities.
If high engagement on a post
can contribute to higher visibility,
then it is important to understand how to increase meaningful engagement with content related to security and privacy.

Prior work has studied what properties of posts about security and privacy on
the ``Reddit'' platform
were correlated with a higher engagement with posts (measured
by the number of comments on a post or a posts's score)~\cite{bhagavatula2021measuring}.
For example, a post that expressed a positive sentiment (e.g., with constructive guidance)
garnered more engagement. This finding was also supported by prior work
that found breach notifications that spoke about the breach constructively encouraged people to learn more about the breach~\cite{bhagavatula2021breach}. The Reddit study also noted
that visual attributes in a post (e.g., emojis or images) encouraged more engagement. People that share advice or information about cybersecurity and privacy on social media
can apply these recommendations by making the posts visually appealing
and by making the posts constructive with a positive sentiment (perhaps by avoiding fear-mongering).

Social networks have also introduced the concept of influential nodes, in the form
of social media influencers~\cite{harrigan2021identifying, subbian2013social}. It is worth exploring if the spread of security and privacy guidance could be increased
if the information is shared by influencers. It is likely that influencers that have a following well beyond the technical sphere will be able to reach wide audiences.

Finally, AI in recommender systems can be effective here. As of not too long ago, recommendation systems in ``Twitter'' factored in a users' social connections to curate that feed. However, Twitter has recently modified this dependency and planned to broaden the content users see across the platform regardless of their connections~\cite{twitter-broader} (``Facebook'' and ``Instagram'' as well~\cite{fb-broader}). While this may still exhibit the issue of predisposed interests, if these algorithms recommend content more broadly without relying heavily on interest, security and privacy guidance has the potential to reach a wider audience.


\subsection{Making security guidance more attractive}

Once the information reaches wider audiences as above,
users need to be willing to accept the cybersecurity advice as applicable to them and important to implement.
This is an additional barrier preventing people from implementing better security and privacy practices.
This barrier is supported by prior work's findings
that people tend to reject security advice if they feel the cost of implementing the advice outweighs the harm if they do not~\cite{good2005stopping, herley2009so}. Additionally, on a cursory analysis, we found some terms used in posts from Table ~\ref{tab:topics} which might exacerbate this barrier, for example, acronyms or industry-specific terms such as ``infosec'' or ``ransomware''.

Security and privacy advice has the potential to be overwhelming,
particularly
when a user is presented with a lot of guidance all at once.
It's also possible that users may find the security concerns they need to be aware
of insurmountable. To that end, security and privacy guidance may be more attractive if it is presented in a minimal way or with minimal jargon.
For example, guidance may often contain several bullet points of instructions. However, often, a little can go a long way and users may be more receptive if they only have to do that ``little'' to be on their way to better security. Additionally, prior work suggests that security and privacy nudges are effective~\cite{acquisti2017nudges}. Notifications
about important security settings can be built into the basic functioning of applications so
that users do not only have to rely on advice they come across themselves.
Given the influence of experts on users' security and privacy decisions~\cite{emami2018influence}, if these notifications include expert recommendations on the decisions for security settings they can take, they may be more likely
to implement these decisions.

\subsection{Relevance of studying ``X''/``Twitter'' data}
Our data only represents the data from one social network. After ``Twitter'' became ``X'', data collection from the platform is tremendously limited and ``X'' may now be rarely used for research.
However, at the time of data collection, ``Twitter'' was considered a meaningful and prominent social network to analyze~\cite{benigni2017isis, yao2021construction}. It is also possible that since this change, usage of the platform has changed in a way that could affect the network structure.
In light of these concerns, we believe that since most social networks still have the potential to contain high homophily, our findings would still be applicable in other popular social networks today (beyond Twitter or X).
\section{Conclusion}
\label{sec:conclusion}



We analyzed data pertaining to $10,159$ users
on Twitter who posted about security and privacy, or quoted, retweeted, or replied to posts about security and privacy to determine whether the information they shared was being circulated in a social media bubble. Specifically, we sought to answer the question of whether security and privacy discussions in social networks occur in communities that would further their reach to a wider audience. Building a follower network graph stemming from those users, we found eight distinct large communities all consisting of users who actively engaged with security- and privacy-related content. We also analyzed the interests of interactors
by looking at the types of posts they generally engaged with on Twitter and found that the general interests of these users already largely aligned with either security and privacy or general technology.

Our findings highlight that discourse around security and privacy is
likely limited to the bubbles surrounding interested users, which is not ideal given the widespread relevance of security and privacy. 
As a result, users that are outside of these bubbles, who may need the exposure more than those in the technology and security sphere, may be less likely to receive this information through social networks. This limitation of such discussions to these bubbles
could potentially be overcome by both increasing the spread of awareness about security and privacy throughout communities in a network, and also by
presenting guidance about security and privacy in a form
that is easy to accept and not overwhelming to implement.

\bibliographystyle{acm}
\bibliography{references}

\appendix
\section{Appendix}
\subsection{Key terms to collect security and privacy posts}
\label{app:key_terms}

Table~\ref{tab:keyterms} contains the final list of key terms we used to collect posts as described in \secref{sec:interactors}.

\begin{table}[hbp]
\begin{tabular}{|p{0.25\linewidth}|p{.75\linewidth}|}
\hline
\textbf{Action} & \textbf{Key terms} \\ \hline
\hline
Keep & confidential computing, cyber attacks, cyberattack, cybersecurity solutions, cybersecurity strategies, cybersecurity threats, hackthebox, network vulnerabilities, phishing, ransomware, surveillance ware, surveillanceware, trusted execution environment                             \\ \hline
Remove & access management, age verification, analytics, authenticate, bots, breach, CDN, Cipher, CISO, compliance, confidential, crypto safety, cyber tasks, data, data collected, data collection, data controllers, data deletion, data enrichment, data governance, data learning, data minimization, data mining, data processors, data protection officers, data quality, data reform, data secured, data use, decrypted, digital attack, digital footprint, digital future, DuckDuckGo\\ \hline
Filter &  access control list, authentication, CAPTCHA, CCPA, chief privacy officer, CISA, cloud security, compromised data, consumer privacy, cyber, cyber attacks, cyber crime, cyber criminals, cyber risk, cyber security regulations, cybercrime, cybersecurity services, data breach, data compromised, data harvesting, data privacy, data privacy law, data privacy management, data privacy policy, data protected, data protection, data protection framework, data protection laws, data protection regulations, data security, ethical hacking, GDPR\\ \hline
\end{tabular}
\caption{A subset of the list of key terms we considered and whether we included the key term in our final set, included it but planned to filter the posts we collected against it, or did not include the key term.}
\label{tab:keyterms}
\end{table}

\subsection{Example security- and privacy-related posts}
\label{sec:example-posts}

Table~\ref{tab:example_posts} contains example posts in our dataset.

\begin{table}[htbp]
\begin{tabular}{|p{\linewidth}|}
\hline
\textbf{Post}  \\ \hline
\hline
Spotify goootttaa be one of my favorite evil data harvesting tech conglomerates
\\
\hline
Changing all my passwords after that virtual webinar!!

Ethical Hacking to the Ministry of Defence

Thanks to [REDACTED] for sharing/scaring more and thanks [REDACTED] for organising

Final evening webinar hosted by me for [REDACTED] 

\#hacking \#apm
\\ \hline \
Difference between Network Access Control List and Security Group in AWS
\#AWS \#CloudComputing  \#programmer \#programming \#dotnetforall \#100DaysOfCode \#DEVCommunity  https://t.co/cYMbgS2k8w
\\ \hline


 Colorado Springs Utilities experiences data breach, customer data compromised https://t.co/YdVHYQO2kY
\\ \hline

Healthcare Provider Exposed Transplant Donor and \#infosec \#infosecurity \#cybersecurity \#threatintel \#threatintelligence \#hacking \#cybernews \#cyberattack \#threathunting 
\#cloudsecurity \#appsec \#malware \#ransomware \#devops \#dfir \#bitcoin \#CISA \#owasp 
https://t.co/Hu1tkKYF4o
\\ \hline
\end{tabular}
\caption{Some examples of posts related to security or privacy in our dataset. We replaced usernames or names with ``[REDACTED]''.}
\label{tab:example_posts}
\end{table}

\let\cleardoublepage=\clearpage

\end{document}